\documentstyle[12pt,epsfig]{article}

\input{tcilatex}

\begin{document}

\baselineskip=15.5pt \pagestyle{plain} \setcounter{page}{1}

\begin{center}

\vskip 3.7 cm

{\LARGE {\bf {Prelogarithmic operators and Jordan blocks in $SL(2)_k$ affine algebra}}}

\vskip 2.1 cm

{\large Gast\'on Giribet}

\vskip 1.0 cm

Instituto de Astronom\'{\i}a y F\'{\i}sica del Espacio, IAFE

C.C. 67 - Suc. 28, (1428) Buenos Aires, Argentina

and

Physics Department, University of Buenos Aires, UBA

gaston@iafe.uba.ar

\vskip 2.1 cm

{\bf Abstract}

The free field description of logarithmic and prelogarithmic operators in
non compact Wess-Zumino-Witten model is analysed. We study the structure of 
the Jordan blocks of the $\widehat{SL(2)_k}$ affine algebra and the role of the
puncture operator in the theory in relation with the unitarity bound.

\end{center}

\section{Introduction}

Wess-Zumino-Witten model formulated on $SL(2)$ manifold is a subject of great
importance for several reasons. Since this model describes the string
dynamics on three dimensional Anti-de Sitter spacetime, its study has
received a renewed interest within the context of $AdS/CFT$ correspondence
conjecture \cite{gks},\cite{kuse}. On the other hand, the $SL(2)$ $WZW$
represents a particular case of non compact conformal field theory and establishes
the discussion about the unitarity and the spectrum of string theory
in presence of nontrivial background fields \cite{unitarity}-\cite{ba}.

More recently, the $SL(2)$ $WZW$ model as an example of
logarithmic conformal field theory has been proposed. Indeed, in reference \cite{kole}-\cite{koled} this $%
CFT$ was studied within this context, and in reference \cite{nisa} some
exact solutions of the Knizhnik-Zamolodchikov equation exhibiting
logarithmic behaviour were found.

Moreover, in \cite{le} it was asserted that logarithmic operators on the
boundary of $AdS_3$ are related to operators on the $WZW$ bulk theory which
are in indecomposable representations of $\widehat{SL(2)}_k$ affine algebra
and several aspects of the logarithmic structure have been analysed in
terms of the $AdS_3/CFT_2$ correspondence \cite{le}-\cite{kowh}.

In this paper we are mainly interested in the free field description of
logarithmic and prelogarithmic operators in the $WZW$ model formulated on
non compact $SL(2)$ manifolds with the intention to present a useful tool to
analyse the identity and origin of this logarithmic structure. In section 2
we review the Wakimoto representation of the conformal field theory in terms
of free fields. In section 3 we analyse the logarithmic operators $\tilde
\Phi _j$ introduced in references \cite{le2} and we obtain the free field
representation of the puncture operator of \cite{kole} which is shown to be
closed related with the primary field $\Phi _{-\frac 12}$; in this section
we also study the Jordan blocks structure of the $\widehat{SL(2)}_k$
Kac-Moody algebra, and some details about the near bounday limit are
commented in terms of the indecomposable representations found previously.
In section 4 we summarize the results and suggest the connection between the
logarithmic structure of $SL(2)$ $WZW$ model and other interesting related
subjects.

\section{The conformal field theory}

\subsection{Wess-Zumino-Witten action on $SL(2,C)/SU(2)$}

Let us begin with a review of the free field realization of the theory.

By using the Gauss parametrization to describe the group elements, the
Wess-Zumino-Witten model formulated on $SL(2,C)/SU(2)$ is given by

\begin{equation}
S=k\int d^2z[\partial \phi \bar \partial \phi +\bar \partial \gamma \partial
\bar \gamma e^{2\phi }]  \label{s1}
\end{equation}

This action describes strings propagating in three dimensional euclidean Anti-de
Sitter space with curvature $-\frac 2k$, metric 
\begin{equation}
ds^2=kd\phi ^2+ke^{2\phi }d\gamma d\bar \gamma  \label{gik}
\end{equation}
and background NS-NS field 
\begin{equation}
B=ke^{2\phi }d\gamma \wedge d\bar \gamma  \label{bik}
\end{equation}

The boundary of $AdS_3$ is located at $\phi =\infty $. Near this region
quantum effects can be treated perturbatively, the exponent in the last term
in (\ref{s1}) is renormalized and a linear dilaton in $\phi $ is generated.
Adding auxiliary fields $(\beta ,\bar \beta )$ and rescaling, the action
becomes 
\begin{equation}
S=\frac 1{4\pi }\int d^2z[\partial \phi \bar \partial \phi -\sqrt{\frac 2{k-2}}R\phi +\beta \bar \partial \gamma +\bar \beta \partial \bar \gamma
-\beta \bar \beta e^{-\sqrt{\frac 2{k-2}}\phi }]  \label{s2}
\end{equation}

This theory has a non compact $SL(2)\times \overline{SL(2)}$ symmetry
generated by currents $J^a(z)$ and $\bar J^a(\bar z)$. In the following we
discuss the holomorphic part of the theory because the same considerations
apply to the antiholomorphic part.

The $SL(2,R)$ current algebra can be expressed in terms of the fields $%
(\phi ,\beta ,\gamma )$ using the Wakimoto representation 
\begin{eqnarray}
J^{+}(z) &=&\beta (z)  \label{uaquimotito} \\
J^3(z) &=&-\beta (z)\gamma (z)-\sqrt{\frac{k-2}2}\partial \phi (z)  \nonumber
\\
J^{-}(z) &=&\beta (z)\gamma ^2(z)+\sqrt{2k-4}\gamma (z)\partial \phi
(z)+k\partial \gamma (z)  \nonumber
\end{eqnarray}

Indeed, considering the free field propagators 
\begin{equation}
\left\langle \phi (z)\phi (w)\right\rangle =-\log (z-w)  \label{corelatuno}
\end{equation}
\begin{equation}
\left\langle \gamma (z)\beta (w)\right\rangle =-\frac 1{(z-w)}
\label{corelatdos}
\end{equation}
it is easy to verify that the OPE of these currents satisfy a $SL(2,R)$
level $k$ Kac-Moody algebra, namely 
\begin{eqnarray}
J^{+}(z)J^{-}(w) &=&\frac k{(z-w)^2}-\frac 2{(z-w)}J^3(w)+...  \label{olgaz}
\\
J^3(z)J^{\pm}(w) &=&\pm\frac 1{(z-w)}J^{\pm}(w)+...  \nonumber \\
J^3(z)J^3(w) &=&\frac{-k/2}{(z-w)^2}+...  \nonumber
\end{eqnarray}

The Sugawara construction leads to the following energy-momentum tensor 
\begin{equation}
T_{SL(2,R)}=\beta \partial \gamma -\frac 12(\partial \phi )^2-\frac 1{\sqrt{%
2(k-2)}}\partial ^2\phi  \label{jemmm}
\end{equation}
and hence the central charge of the theory is 
\begin{equation}
c=3+\frac{6}{k-2}  \label{charche}
\end{equation}

The spectrum of the theory can be obtained analysing the $SL(2,R)$ Kac-Moody
primary states. They are labeled by two quantum numbers $\left|
j,m\right\rangle $, where

\begin{eqnarray}
\hat C\left| j,m\right\rangle &=&j(j+1)\left| j,m\right\rangle
\label{guinpa} \\
J_0^3\left| j,m\right\rangle &=&m\left| j,m\right\rangle  \nonumber
\end{eqnarray}
and $\hat C=-\frac 12(J_0^{+}J_0^{-}+J_0^{-}J_0^{+})+J_0^3J_0^3$ is the
Casimir (the subindices refer to the zero-modes of the currents). Notice
that the Casimir value is invariant under $j\rightarrow -j-1$ and therefore
one can consider only states with $j>-{\frac 12}$.

The Kac-Moody primarie states $\left| j,m\right\rangle $ are defined by the
condition 
\begin{eqnarray}
J_n^a\left| j,m\right\rangle =0,\quad n>0  \label{wimpa}
\end{eqnarray}
while the descendents are constructed by acting with $J_{-n}^a$, $n>0$ on (%
\ref{wimpa}).

In terms of the quantum numbers $j$ and $m$, the vertex operators create
states from the $SL(2)$ vaccum, namely 
\[
\lim_{z\rightarrow 0}\Phi _{j,m}\left| 0\right\rangle =\left|
j,m\right\rangle 
\]

These operators have the conformal dimension $h(j)$ given by 
\[
h(j)=-\frac 2{\alpha _{+}^2}j(j+1) 
\]
where $\alpha _{+}^2=2(k-2)$.

\subsection{Near boundary limit of the vertex operator}

It is usual to introduce auxiliary coordinates ($x,\bar x$) in order to
organize the $SL(2)$ representations. In this picture, the vertex operators
are given by \cite{te} 
\begin{equation}
\Phi _j(x,\bar x)=\frac 1\pi \left( \left| \gamma -x\right| ^2e^{\frac
1{\alpha _{+}}\phi }+e^{-\frac 1{\alpha _{+}}\phi }\right) ^{-2j-2}
\label{ffeo}
\end{equation}
and are related with the picture $\Phi _{j,m,\bar m}$ by a Fourier
transform defined as follows 
\begin{equation}
\Phi _{j,m,\bar m}=\int d^2x\Phi _j(x,\bar x)x^{j-m}\bar x^{j-\bar m}
\label{furier}
\end{equation}

In this paper we are mainly interested in the situation when $\phi $ is
large since the free field aproximation is trusted in that case; thus it
will be useful to take the large $\phi $ limit of the expression (\ref{ffeo}%
) to obtain the form of the vertex operators in the near boundary region.

Taking into account the following representation of the two-dimensional
Dirac function 
\[
\delta ^{(2)}(\gamma -x)=\delta (\gamma -x)\delta (\bar \gamma -\bar x)=%
\frac{n-1}\pi \lim_{\varepsilon \rightarrow 0}\frac{\varepsilon ^{2n-2}}{%
\left( \varepsilon ^2+\left| \gamma -x\right| ^2\right) ^n} 
\]
, we can relate $n=2j+2$ and $\varepsilon =e^{-\frac \phi {\alpha _{+}}}$ in
order to write 
\[
\lim_{\phi \rightarrow \infty }\Phi _j(x,\bar x)=\frac{e^{\frac{2\phi }{%
\alpha _{+}}j}}{2j+1}\delta ^{(2)}(\gamma -x)+others 
\]

This $\delta $-term is only dominant in the region $\gamma \approx x$; and
the complete expression for the leading behaviour in $\phi $ is given by 
\cite{kuse} 
\begin{equation}
\Phi _j(x,\bar x)\approx \frac{e^{\frac{2j}{\alpha _{+}}\phi }}{2j+1}\delta
^{(2)}(\gamma -x)+\frac 1\pi \left| \gamma -x\right| ^{-4j-4}e^{-\frac{2j+2}{%
\alpha _{+}}\phi }+O(e^{-\frac{2(j+2)}{\alpha _{+}}\phi })
\label{cachabacha}
\end{equation}

In the particular case $j=-\frac 12$ both $e^{\frac{2j}{\alpha _{+}}\phi }$
and $e^{-\frac{2j+2}{\alpha _{+}}\phi }$ are the dominant terms and we are
in presence of a \emph{resonance}. Indeed, it will be very useful to our
further analysis to sudy the expression of the leading term in the large $%
\phi $ expansion in that particular case. In order to do that, we can start from the
continuous representation $j=-\frac 12+i\varepsilon $ and then take the
limit $\varepsilon \rightarrow 0$.

Thus, for large values of $\phi $ we have 
\[
\Phi _{-\frac 12+i\varepsilon }(x)\approx \frac{e^{\frac{-1+2i\varepsilon }{%
\alpha _{+}}\phi }}{2i\varepsilon }\delta ^{(2)}(\gamma -x)+\frac 1\pi
\left| \gamma -x\right| ^{-2-4i\varepsilon }e^{\frac{-1-2i\varepsilon }{%
\alpha _{+}}\phi }+O(e^{-\frac 1{2\alpha _{+}}\phi })= 
\]
\begin{equation}
=\left( 1+\frac{2i\varepsilon }{\alpha _{+}}\phi +O(\varepsilon
^2\phi ^2)\right) \times \frac 1{2i\varepsilon }e^{-\frac \phi {\alpha
_{+}}}\delta ^{(2)}(\gamma -x)+\frac 1\pi \left| \gamma -x\right|
^{-2-4i\varepsilon }e^{-\frac{1+2i\varepsilon }{\alpha _{+}}\phi }
\end{equation}

Now, it is possible to perform the Fourier transform (\ref{furier}) to
obtain the ($m,\bar m$) picture of this operator. And it is very interesting
to notice that the term $\frac 1{2i\varepsilon }$ is cancelated with a term $%
-\frac 1{2i\varepsilon }$ coming from the Fourier transform of $\frac 1\pi
\left| \gamma -x\right| ^{-2}$. Indeed, by using that 
\begin{equation}
\int d^2x\left| x\right| ^{-2r}\left| 1-x\right| ^{-2s}=\pi \frac{\Gamma
(1-r)\Gamma (1-s)\Gamma (r+s-1)}{\Gamma (r)\Gamma (s)\Gamma (2-r-s)}
\label{laintegral}
\end{equation}
we can write 
\[
\Phi _{-\frac 12+i\varepsilon ,m,\bar m}\approx \left( \frac 1{2i\varepsilon
}+\frac \phi {\alpha _{+}}\right) \gamma ^{-\frac 12+i\varepsilon -m}\bar
\gamma ^{-\frac 12+i\varepsilon -\bar m}e^{-\frac \phi {\alpha _{+}}}+ 
\]
\[
+\left( -\frac 1{2i\varepsilon }-\zeta \right) \left( 1- \frac {2i\varepsilon \phi} {\alpha _{+}}\right) \times
\]
\begin{equation}
\times \frac{\Gamma (\frac
12+m+i\varepsilon )\Gamma (\frac 12-\bar m+i\varepsilon )}{\Gamma
(1+2i\varepsilon )\Gamma (\frac 12+m-i\varepsilon )\Gamma (\frac 12-\bar
m-i\varepsilon )}\gamma ^{-\frac 12-i\varepsilon -m}\bar \gamma ^{-\frac
12-i\varepsilon -\bar m}e^{-\frac \phi {\alpha _{+}}}+O%
(\varepsilon )  \label{odv}
\end{equation}
where $\zeta $ is the Euler-Mascheroni constant. And taking the limit $%
\varepsilon $ going to cero, we obtain 
\begin{equation}
\Phi _{-\frac 12,m,\bar m}=\left( \frac {2\phi} {\alpha _{+}}+\ln (\gamma
)-\zeta \right)
\gamma ^{-\frac 12-m}\bar \gamma ^{-\frac 12-\bar m}e^{-\frac \phi {\alpha
_{+}}}  \label{ottau}
\end{equation}

The above expression has to be understood as the free field representation
of the vertex operator for the particular case $j=-\frac 12$. Actually, this
free field description of $\Phi _{-\frac 12}$ completes the answer to the
suggestion effectuated in reference \cite{kole} about the existence of
operators with the form $\Phi \sim \phi e^{-\frac \phi {\alpha _{+}}}$ in
the $WZW$ model on non-compact groups.

It could be interesting to mention that a free field description for the
puncture operator can be done in terms of the Dotsenko conjugate
representation \cite{dotsenko} (see also \cite{dotsenko2, kogancero,
gastoncarmen3}). For instance, it is possible to demostrate that the
functional form for such conjugated vertex operator in the limit $%
j\rightarrow -\frac 12$ can be given by $\left( \ln \beta -\frac 2{\alpha
_{+}}\phi \right) \beta ^{k-2}e^{\frac{(3-2k)}{\alpha _{+}}\phi }$; notice
that this field is, in fact, a primary field.

It is also important to comment that if quantum (finite-$k$) effects are taken
into account \cite{te} the form for the vertex operator to be considered
becomes

\[
\Phi _{-\frac 12+i\varepsilon }(x)\approx \frac{e^{\frac{-1+2i\varepsilon }{%
\alpha _{+}}\phi }}{2i\varepsilon }\delta ^{(2)}(\gamma -x)+R(j)\left|
\gamma -x\right| ^{-2-4i\varepsilon }e^{\frac{-1-2i\varepsilon }{\alpha _{+}}%
\phi }+O(e^{-\frac 1{2\alpha _{+}}\phi }) 
\]
where the reflection factor $R(j)$ is now given by 
\begin{equation}
R_{\left( j=-\frac 12+i\varepsilon \right) }=\frac 1\pi \left( \pi \frac{%
\Gamma (1-\frac 1{k-2})}{\Gamma (1+\frac 1{k-2})}\right) ^{2i\varepsilon }%
\frac{\Gamma (1-\frac{2i\varepsilon }{k-2})}{\Gamma (1+\frac{2i\varepsilon }{%
k-2})}
\end{equation}

And, in this case, we can see from this new expression that the result (\ref
{ottau}) is recovered in the limit $\varepsilon \rightarrow 0$.

On the other hand, it is straightforward to see from (\ref{cachabacha}) that
the ($m,\bar m$) picture for the case $j>-\frac 12$ is given by the usual
expression 
\begin{equation}
\Phi _{j>-\frac 12,m,\bar m}=\gamma ^{j-m}\bar \gamma ^{j-\bar m}e^{\frac{%
^{2j}}{\alpha _{+}}\phi }  \label{odeve}
\end{equation}
up to a factor $2j+1>0.$

It will be convenient for our analysis to introduce new notation to denote
the fields of the theory, being 
\begin{eqnarray}
\Psi _{j,m,\bar m} &\equiv &\frac 2{\alpha _{+}}\phi \gamma ^{j-m}\bar
\gamma ^{j-\bar m}e^{\frac{2j}{\alpha _{+}}\phi }  \nonumber  \label{faaa} \\
\Theta _{j,m,\bar m} &\equiv &\gamma ^{j-m}\bar \gamma ^{j-\bar m}e^{\frac{%
^{2j}}{\alpha _{+}}\phi }  \label{faaa}
\end{eqnarray}
and hence, we can write $\Phi _{j>-\frac 12,m,\bar m}=\Theta _{j,m,\bar m}$
and $\Phi _{-\frac 12,m,\bar m} \sim \Psi _{-\frac 12,m,\bar m}$. Notice that we have taken into account only the leading term $\sim \phi e^{\frac{2j}{\alpha _{+}}\phi}$ in this definition of nomenclature.

\section{Logarithmic and prelogarithmic operators}

\subsection{Logarithmic conformal field theory}

A logarithmic conformal field theory is a $CFT$ with not diagonalizable $L_0$
Virasoro generator. In these theories, we have ordinary primaries operators $%
\Phi $ as well as logarithmic operators $\tilde \Phi $ which form Jordan blocks in
Virasoro algebra with the following structure \cite{gu} 
\begin{eqnarray}
T(z)\Phi (w) &=&\frac{h\Phi }{(z-w)^2}+\frac{\partial \Phi }{(z-w)}+... 
\nonumber  \label{logchi} \\
T(z)\tilde \Phi (w) &=&\frac{h\tilde \Phi }{(z-w)^2}+\frac{\xi \Phi }{(z-w)^2%
}+\frac{\partial \tilde \Phi }{(z-w)}+...  \label{logchi}
\end{eqnarray}
where $\xi $ is a complex number.

Recently, the logarithmic conformal field theories have been extensively
studied in realtion with several topics in theoretical physics (see \cite
{morosa}); and as we have commented before, it has been proposed that the $%
SL(2)$ $WZW$ model could be an example of this class of $CFT$.

Indeed, by exploring the similarity between the non-compact $WZW$ and the $%
c_{q,p}$ models, references \cite{kole} and \cite{le2} have mentioned the possibility to
define the logarithmic operators in the $SL(2)$ $WZW$ models as 
\begin{equation}
\tilde \Phi _j(x)=\frac d{dj}\Phi _j(x)=\frac 1\pi \left( \left| \gamma
-x\right| ^2e^{\frac \phi {\alpha _{+}}}+e^{-\frac \phi {\alpha
_{+}}}\right) ^{-2(j+1)}\ln \left( \left| \gamma -x\right| ^2e^{\frac \phi
{\alpha _{+}}}+e^{-\frac \phi {\alpha _{+}}}\right) ^2  \label{ol}
\end{equation}
being $\Phi _j(x)$ the ordinary primary field (\ref{ffeo}). This operator
has the following conformal structure 
\begin{equation}
L_0\tilde \Phi _j(z,x)=-\frac{j(j+1)}{k-2}\tilde \Phi _j(z,x)-\frac{2j+1}{k-2%
}\Phi _j(z,x)  \label{qqqq}
\end{equation}
where we can clearly see that $\tilde \Phi _{-\frac 12}$ appears as a
particular case, being actually an ordinary primary state while $\tilde \Phi
_{j\neq -\frac 12}$ presents the quoted logarithmic structure \cite{le2}.
This class of operators belonging to a logarithmic branch but with the
property to be a primary field of the Virasoro algebra are known as
prelogarithmic operators (or \emph{puncture operators} in the Liouville
nomenclature).

It was explicitly shown in \cite{kole} that the prelogarithmic operator can
be in a indescomposable representation of the Kac-Moody algebra even though
they are primary fields of the Virasoro algebra.

Now, let us to analyse the free field description of the $\tilde \Phi _{j,m}$
operators. To do that, we can derivate the expression (\ref{odv}) with
respect to $j$ or directly take the large $\phi $ limit in (\ref{ol}) as it
was done for the operator (\ref{ffeo}). In both cases, the result is given
by 
\begin{equation}
\tilde \Phi _{j,m}=\gamma ^{j-m}e^{\frac{2j}{\alpha _{+}}\phi }(\ln (\gamma
)+\frac 2{\alpha _{+}}\phi )=\Theta _{j,m}(\ln (\gamma )+\frac 2{\alpha
_{+}}\phi )  \label{dalee}
\end{equation}
And now, by using the prescription (see \cite{ra} and references therein) 
\begin{equation}
\left\langle \beta (z)f(\gamma (w))\right\rangle =:\left( \beta (z)+\frac
1{(z-w)}\frac \partial {\partial \gamma }\right) f(\gamma (w)):+...
\label{rass}
\end{equation}
it is easy to verify from (\ref{jemmm}) that the $\tilde \Phi _{j,m}$ fields
exhibit the Jordan block structure of Virasoro algebra given by (\ref{qqqq}%
), namely 
\[
T_{SL(2,R)}(z)\tilde \Phi _{j,m}(w)=\left( -\frac{j(j+1)}{k-2}\right) \frac{%
\tilde \Phi _{j,m}}{(z-w)^2}-\left( \frac{2j+1}{k-2}\right) \frac{\Phi _{j,m}%
}{(z-w)^2}+\frac{\partial \tilde \Phi _{j,m}}{(z-w)}+... 
\]

The fact that the ($m,\bar m$) picture exhibits a similar logarithmic
structure to the ($x,\bar x$) picture becomes important if it is taken into
account that it is general not clear the subtle detail of the correspondence 
$\Phi _{j,m,\bar m}\longleftrightarrow \Phi _j(x,\bar x),$ as it was
remarked in reference \cite{kuse}.

\subsection{The puncture operator and the resonance}

Let us present an heuristic argument within the context of logarithmic structure of the operator product expansion.
As it was mentioned in the last subsection, the particular case $\tilde \Phi
_{-\frac 12,m}$ is actually a primary field of Virasoro algebra. This is the
puncture operator of the theory and it was extensively studied in reference 
\cite{le2}.

On the other hand, since it is our intention to make clear the
relation between the operator $\tilde \Phi _{-\frac 12}$ and the
prelogarithmic field analysed in reference \cite{kole}, it is important
to remark that the operators $\Psi _{j,m}$ also exhibit the logarithmic
structure (\ref{qqqq}). This fact could induce to interpret the $\Psi
_{-\frac 12,m}$ as a puncture operator of the theory, and we have shown
before that this field is actually the large $\phi $ limit of $\Phi _{-\frac
12,m}$. This is a very important point, the free field representation of the
primary field with $j=-\frac 12$ presents the functional form of the
logarithmic operators $\sim \phi e^{\frac{2j}{\alpha _{+}}\phi }.$ In fact,
before expanding it in power of $\phi $ and performing the Fourier transform, the
aspect of the $\Phi _{-\frac 12}(x)$ was substantially different from $%
\tilde \Phi _{-\frac 12}(x),$ which shows that the free field description in
terms of $\Phi _{j,m}$ suggestes more explicitly the particular connection
existing between $\tilde \Phi _{-\frac 12}$ and $\Phi _{-\frac 12}$.

In \cite{le2} it was commented that it is not possible to have $\Phi
_{-\frac 12}$ in the spectrum of the $SL(2)$ $WZW$ model without $\tilde
\Phi _{-\frac 12}$ also been included. Thus, in order to analyse the
possibility of the occurence of logarithmic operators in the spectrum of the 
$SL(2)$ model, the operator product expansion of two fields have been
studied in the literature (see \cite{le2} and \cite{cf}). Indeed, taking
into account the OPE's 
\begin{equation}
\Theta _{j_1,m_1}(z)\Theta _{j_2,m_2}(w)\sim \left| z-w\right| ^{-\frac
4{k-2}j_1j_2}\Theta _{j_1+j_2,m_1+m_2}(w)+...  \label{laotran}
\end{equation}
and 
\begin{eqnarray}
\Theta _{j_1,m_1}(z)\Psi _{j_2,m_2}(w) &\sim &\left| z-w\right| ^{-\frac
4{k-2}j_1j_2}(\Psi _{j_1+j_2,m_1+m_2}(w)+  \nonumber  \label{ochete} \\
&&-\frac{2j_1}{k-2}\ln \left| z-w\right| \Theta _{j_1+j_2,m_1+m_2}(w))+...
\label{ochete}
\end{eqnarray}
it is possible to observe that the logarithm in the operator product
expansion manifestly appears.

On the other hand, it is very important to notice that the expression (\ref
{ochete}) explicitly shows the mixing terms raising from the presence of a
field with the form $\Phi _{-\frac 12,m}\sim \Psi _{-\frac 12,m}$ since that
a logarithmic field $\Psi _{j\neq -\frac 12,m}$ appears in the operator
product expansion if $j_1\neq 0$.

Actually, the logarithmic contributions also appear when the operator
product of the fields $\tilde \Phi _j$ are considered, namely

\begin{equation}
\Theta _{j_1,m_1}(z)\tilde \Phi _{j_2,m_2}(w)\sim \left| z-w\right| ^{-\frac
4{k-2}j_1j_2}(\tilde \Phi _{j_1+j_2,m_1+m_2}(w)-\frac{2j_1}{k-2}\ln \left|
z-w\right| \Theta _{j_1+j_2,m_1+m_2}(w))+...  \label{laotran}
\end{equation}

Now, we would like to emphasize the differences and similarities between the
operators $\Psi _{-\frac 12,m}$ and $\tilde \Phi _{-\frac 12,m}$ since these
fields play a crucial role in the study of logarithmic structure of $SL(2)$ $%
WZW$ model \cite{le2}. It is clear that the linear term in $\phi $ that it
is present in both (\ref{odv}) and (\ref{dalee}) dominates near the boundary
and thus these fields have the same large $\phi $ behaviour; however we will
see in the next subsection that the presence of the $\ln (\gamma )$ in (\ref
{dalee}) makes that the structure of the corresponding Jordan blocks of
Kac-Moody algebra are quite different. This fact shows us that the leading
terms in the large $\phi $ expansion are in general not sufficient to
analyse the logarithmic structure of the current algebra, and then it would
be necessary not to neglect the $\ln (\gamma )$ term in order to study it.

\subsection{Jordan blocks of $\widehat{SL(2)}_k$ affine algebra}

It was shown in reference \cite{kole} that the fields $\Psi _{j,m}$ form
Jordan blocks in the Kac-Moody $\widehat{SL(2)}_k$ affine algebra even
though $j=-\frac 12.$ This logarithmic structure of the current algebra is
given by 
\begin{eqnarray}
J^{+}(z)\left( \frac{\alpha _{+}}2\Psi _{j,m}(w)\right) &=&\frac{j-m}{(z-w)}%
\left( \frac{\alpha _{+}}2\Psi _{j,m+1}\right) +...  \nonumber  \label{kagan}
\\
J^3(z)\left( \frac{\alpha _{+}}2\Psi _{j,m}(w)\right) &=&\frac
m{(z-w)}\left( \frac{\alpha _{+}}2\Psi _{j,m}\right) +\frac{\alpha _{+}/2}{%
(z-w)}\Theta _{j,m}+...  \nonumber \\
J^{-}(z)\left( \frac{\alpha _{+}}2\Psi _{j,m}(w)\right) &=&-\frac{j+m}{(z-w)}%
\left( \frac{\alpha _{+}}2\Psi _{j,m-1}\right) -\frac{\alpha _{+}}{(z-w)}%
\Theta _{j,m-1}+...  \label{kagan}
\end{eqnarray}

On the other hand, as it was commented in the previous subsection, from (\ref
{dalee}) it is easy to verify that the operators $\tilde \Phi _{j,m}$ form a
different and \emph{more symmetric} Jordan block, namely 
\begin{eqnarray}
J^{\pm }(z)\tilde \Phi _{j,m}(w) &=&\frac{(\pm j-m)}{(z-w)}\tilde \Phi
_{j,m\pm 1}(w)\pm \frac 1{(z-w)}\Theta _{j,m\pm 1}(w)+...  \nonumber \\
J^3(z)\tilde \Phi _{j,m}(w) &=&\frac m{(z-w)}\tilde \Phi _{j,m}(w)+...
\label{mejor}
\end{eqnarray}

Thus, the nondiagonal structure (\ref{kagan}) must be understood as the
Jordan blocks of the large $\phi $ limit of the primary field $\Phi _{-\frac
12}$ rather than the corresponding to the field $\tilde \Phi _j=\frac
d{dj}\Phi _j$. Then, we have obtained in (\ref{mejor}) a new Jordan cell of
the affine $\widehat{SL(2)}_k$ algebra. It gives us an example to show
how the structure of the Jordan blocks of the current algebra depends on the
functional form of the field representation beyond the leading terms in
large $\phi $ expansion.

Notice that in the $\alpha _{+}\rightarrow 0$ limit \.($i.e.\ k\rightarrow
2^{+}$) the Jordan structure of the operators $\frac{\alpha _{+}}2\Psi
_{j,m} $ becomes diagonal, while the blocks for the fields $\tilde \Phi
_{j,m}$ remain in its Jordan form in that limit. This fact remarks the
differences between (\ref{kagan}) and (\ref{mejor}).

If, on the other hand, we consider the operator defined as 
\begin{equation}
\Xi _{j,m}\equiv \tilde \Phi _{j,m}-\Psi _{j,m}=\ln (\gamma )\Theta _{j,m}
\label{indio}
\end{equation}
it is possible to obtain the following nondiagonal realization 
\begin{eqnarray}
J^{\pm }(z)\Xi _{j,m}(w) &=&\frac{(\pm j-m)}{(z-w)}\Xi _{j,m\pm 1}(w)+\frac
1{(z-w)}\Theta _{j,m\pm 1}(w)+...  \nonumber  \label{kaganto} \\
J^3(z)\Xi _{j,m}(w) &=&\frac m{(z-w)}\Xi _{j,m}(w)-\frac 1{(z-w)}\Theta
_{j,m}(w)+...  \label{kaganto}
\end{eqnarray}
where (\ref{indio}) is a primary field according to the prescription (\ref{rass}).
Then, it would be a new realization of a primary field in Virasoro algebra
but bellonging to an indecomposable representation of Kac-Moody algebra.
These new operators $\Xi _{j,m}$ form an infinite set of primary states in
the Virasoro algebra with logarithmic structure in the $\widehat{SL(2)}_k$
affine Kac-Moody algebra. And, on the other hand, it is possible to verify
that the operator product expansion of two of these fields does not contain
mixing terms with prelogarithmic fields.

\subsection{Spacetime stress tensor on a long string}

Let us comment in this subsection an important point within the context
of $AdS_3/CFT_2$ correspondence.

In a recent paper \cite{d1d5} Seiberg and Witten have obtained the
spacetime stress tensor of the theory on the single long string solution.
This stress tensor $T_{tot}$ is obtained by twisting the worldsheet tensor (%
\ref{jemmm}) to 
\begin{equation}
T_{tot}=T_{SL(2,R)}-\partial J^3  \label{sewi}
\end{equation}

In fact, this result leads to compute the Brown-Henneaux central charge
and the corresponding conformal dimension for the primary spacetime
fields.

Since it was analysed in \cite{kole} that the logarithmic structure can be
altered in models with modified stress tensor, it would be interesting to
investigate the logarithmic behaviour of the fields (\ref{ottau}) and (\ref
{dalee}) in terms of the theory defined by (\ref{sewi}). In order to do
so, we consider the following form for the modified stress tensor 
\begin{equation}
T_{tot}=T_{SL(2,R)}-Q\partial J^3=(1+Q)\beta \partial \gamma +Q\partial
\beta \gamma -\frac 12(\partial \phi )^2+\left( \frac{\alpha _{+}}2Q-\frac
1{\alpha _{+}}\right) \partial ^2\phi  \label{sewis}
\end{equation}
which includes models considered in the study of non compact $CFT$ and
two-dimensional gravity.

Thus, by computing the operator product expansion it is possible to verify
that the structure of the Virasoro algebra in that case is given by the
following expression

\begin{equation}
T_{tot}(z)\Phi _{j,m}(w)=\left( -\frac{j(j+1)}{k-2}+Qm\right) \frac{\Phi
_{j,m}}{(z-w)^2}+\frac{\partial \Phi _{j,m}}{(z-w)}+...  \label{dizza}
\end{equation}
while for the fields $\tilde \Phi _j$ and $\Psi _j$ we obtain 
\[
T_{tot}(z)\tilde \Phi _{j,m}(w)=\left( -\frac{j(j+1)}{k-2}+Qm\right) \frac{%
\tilde \Phi _{j,m}}{(z-w)^2}+\left( \frac{2j+1}{k-2}\right) \frac{\Theta
_{j,m}}{(z-w)^2}+\frac{\partial \tilde \Phi _{j,m}}{(z-w)}+... 
\]
\begin{equation}
T_{tot}(z)\Psi _{j,m}(w)=\left( -\frac{j(j+1)}{k-2}+Qm\right) \frac{\Psi
_{j,m}}{(z-w)^2}+\left( \frac{2j+1}{k-2}+Q\right) \frac{\Theta _{j,m}}{%
(z-w)^2}+\frac{\partial \Psi _{j,m}}{(z-w)}+...  \label{bombao}
\end{equation}
and thus 
\begin{equation}
T_{tot}(z)\Xi _{j,m}(w)=\left( -\frac{j(j+1)}{k-2}+Qm\right) \frac{\Xi _{j,m}%
}{(z-w)^2}-Q\frac{\Theta _{j,m}}{(z-w)^2}+\frac{\partial \Xi _{j,m}}{(z-w)}%
+...  \label{o}
\end{equation}

Indeed, from these expressions we can compare the logarithmic structure of
both theories (\ref{jemmm}) and (\ref{sewi}); and it is possible to observe
that the Jordan blocks of Virasoro algebra for the fields $\tilde \Phi _j$
remain unchanged while for the fields $\Phi _{-\frac 12}$ do not. This fact
makes more evident that the Jordan structure is defined by the different
terms in the large $\phi $ expansion.

Notice that since we have shown the relation (\ref{ottau}) between the
fields $\Psi _{-\frac 12,m}$ and $\Phi _{-\frac 12,m}$, it is possible to
afirm that the leading term in the near boundary limit of the primary field $\Phi _{-\frac 12}$ becomes logarithmic for the
theory defined by (\ref{sewi}). This is in agreement with one of the results
of reference \cite{le2}, where it was concluded that the worldsheet primary
field $\Phi _{-\frac 12}$ is a logarithmic field in the spacetime theory.
It is immediate to see from (\ref{dizza}) and (\ref{bombao}) that the
parimary fields $\Xi _{j,m}$ also become logarithmic in the twisted theory $%
Q\neq 0$.

On the other hand, from the expression (\ref{bombao}) we can see that in the
spacetime theory on the long string $Q=\pm 1,$ as well as in the
world-sheet theory $Q=0$, the logarithmic structure of the Virasoro algebra
of fields $\tilde \Phi _j$ aproximates to the structure of fields $\Psi _j$
in the limit $\alpha _{+}\rightarrow 0$; and this fact is consistent with
the functional forms (\ref{ottau}) and (\ref{odeve}). However, as we have
seen before, the logarithmic structure of Kac-Moody algebra of these fields
remain distinct in that limit.

Notice that in the case $Q=-1$ the prelogarithmic operators are $\tilde \Phi
_j$ with $j=-\frac 12$ and $\Psi _j$ with $j=\frac{k-3}2$, and these are
precisely the minimum and maximum value of $j$ in the discrete serie
according to satisfy the unitarity bound (to consider the case $Q=+1$
instead $Q=-1$ is consistent with the reflection symmetry $j\leftrightarrow
-j-1$). Moreover, $\Psi _{-\frac 12}$ appears as the prelogarithmic operator
for the worldsheet theory whereas $\Psi _{\frac{k-3}2}$ is
the prelogarithmic operator in the theory on a single ($\omega =\pm 1$) long
string solution.

\subsection{Zero mode contribution and logarithmic operators}

Another interesting aspects of the free field representation is the fact
that it leads to draw a rare relation existing between the near boundary
behaviour of the logarithmic operators (\ref{ol}) and the $r$-string states
introduced by Bars, Deliduman and Minic in reference \cite{ba}.

Bars $et$ $al.$ have proposed that the inclusion of the zero mode
contribution plays a crucial role in the study of the spectrum of the $SL(2)$
$WZW$ model. Actually, it was argued that this contribution is important to
describe the winding states in the $AdS_3$ spacetime.

Indeed, the inclusion of the zero mode in refrerence \cite{ba} leads to
generalize the vertex operator (\ref{ffeo}) in order to incorporate the
description of the $\omega $-winding fields $\Phi _j^{m\omega }(x),$ namely 
\[
\Phi _{j=-\frac 12+i\lambda }^{m\omega }(x)=\Phi _{j=-\frac 12+i\lambda
}(x)\times \left( e^{\frac \phi {\alpha _{+}}}\left| \gamma -x\right|
\right) ^{-2(\sqrt{(k-2)m\omega -\lambda ^2}-i\lambda )}\times 
\]
\begin{eqnarray}
&&\times _2F_1(\sqrt{(k-2)m\omega -\lambda ^2}-i\lambda ,\sqrt{(k-2)m\omega
-\lambda ^2}-i\lambda ;1+  \nonumber  \label{dalebars} \\
&&+2\sqrt{(k-2)m\omega -\lambda ^2}-i\lambda ;-e^{-\frac 2{\alpha _{+}}\phi
}\left| \gamma -x\right| ^{-2})  \label{dalebars}
\end{eqnarray}
which coincides with the usual vertex operator $\Phi _j(x)$ in the
particular case $m\omega =0$, namely 
\begin{equation}
\Phi _{j=-\frac 12+i\lambda }^0(x)=\Phi _{j=-\frac 12+i\lambda }(x)
\end{equation}

These $SL(2)$ states are primaries and have a conformal dimension given by 
\begin{equation}
h(j)=-\frac 2{\alpha _{+}^2}j(j+1)-m\omega =\frac 2{\alpha _{+}^2}(\frac
14+\lambda ^2)-m\omega
\end{equation}
with a new $m\omega $ contribution that becomes important within the context
of the unitarity problem \cite{ba},\cite{rs}.

And taking the large $\phi $ limit we can see from (\ref{dalebars}) that the
following behaviour is obtained near the boundary \cite{ba}

\begin{equation}
\Phi _{j=-\frac 12+i\lambda }^{m\omega }(x)\sim N_{(\lambda
,m\omega ,k)}\left( \ln \left| \gamma -x\right| ^2+\frac 2{\alpha _{+}}\phi
\right) \label{wwwwwr}
\end{equation}
where $N_{(\lambda ,m\omega ,k)}=\frac{\Gamma (1+2\sqrt{km\omega
-\lambda ^2})}{\Gamma (\sqrt{km\omega -\lambda ^2}-i\lambda )\Gamma (\sqrt{%
km\omega -\lambda ^2}+1+i\lambda )}$. Hence, from (\ref{dalee}) it is
possible to observe that the large $\phi $ behaviour of (\ref{wwwwwr})
coincide with the power expansion in $\phi $ of the logarithmic operators $%
\tilde \Phi _j(x)$, including a term with $\ln \left| \gamma -x\right| ^2$.

Notice that the presence of terms of the form $\delta ^{(2)}(\gamma -x)$ in
the large $\phi $ expansion of $\Phi _{j=-\frac 12+i\lambda }$ makes it
impossible to neglect the term $\ln \left| \gamma -x\right| ^2$ in general.
Thus, we see from (\ref{wwwwwr}) that the vertex operators introduced in 
\cite{ba} form Jordan bloks of the free field realization of the $SL(2)$
current algebra with the form of the Jordan cells analysed in the previous
sections of this note.

Moreover, taking into account that it is usual to assume that the
logarithmic sectors in string theory are though to generate additional
symmetries \cite{le2}, this fact suggests that it could be interesting to
mention a connection between the zero mode contribution remarked by Bars $et$ $%
al.$, the indecomposable operators $\Xi _{j,m}$ and the logarithmic fields $%
\tilde \Phi _{j,m}$. Actually, this is for futher investigation
(recently, the winding modes have been successfully described in terms of
spectral flow symmetry \cite{winding}).

\section{Conclusions}

We have studied the free field representation of the logarithmic and
prelogarithmic operators in the Wess-Zumino-Witten model formulated on $%
SL(2)$. We have shown that the $\Phi _{j,m,\bar m}$ picture is
useful to analyse the logarithmic structure of the Virasoro and Kac-Moody
algebra.

The similarities and differences existing between the puncture operator of
reference \cite{le2} and the large $\phi $ behaviour of the resonance $%
j=-\frac 12$ were pointed out. We have shown that the leading terms in the
near boundary region of both operators coincide even though the description
in terms of the auxiliary variables ($x,\bar x$) does not manifest this
relation explicitly. This fact shows how the ($m,\bar m$) picture results
useful to analyse the particular connection between $\tilde \Phi _{-\frac
12} $ and $\Phi _{-\frac 12}$ (while the Fourier conjugate picture ($x,\bar
x $) has shown to be useful to explore the logarithmic structure of the
boundary $CFT$ in previous works \cite{le}-\cite{kowh}).

The Jordan blocks of the Kac-Moody algebra were written down for operators
of the form $\tilde \Phi _j=\frac d{dj}\Phi _j$ and $\Phi _{-\frac 12}$.
These results show how the different terms in the large $\phi $ expansion
determinate the logarithmic structure of the current algebra. Actually, we
have argued that the non diagonal structure found in reference \cite{kole}
must be understood as the Jordan blocks of the large $\phi $ limit of the
primary field $\Phi _{-\frac 12}$ rather than the corresponding to the
operators $\frac d{dj}\Phi _j$.

On the other hand, new free field realization of primary operators $\Xi _{j,m}$
belonging to indecomposable representations of Kac-Moody algebra have been
obtained and we have also analysed the operator porduct expansion of two
fields in order to make explicit the occurence of the logarithmic structure
in terms of the free field description.

We have also shown that the primary states of the original theory become
logarithmic when a twisted stress tensor is considered and this fact is
consistent with results of previous works in this matter. The minimum and
maximum values of $j$ admissible by the unitarity bound were seen as
prelogarithmic states for the worldsheet theory and for the spacetime
theory on a single long string solution respectively. 
\[
\]

\textbf{Acknowledgements}

I am very grateful to C. N\'u\~nez for stimulating collaboration. I also
thank A. Nichols and J.G. Russo for useful conversations. This work was
supported by CONICET.

\end{document}